\documentclass[
]{ceurart}

\usepackage{listings}
\usepackage{natbib}
\usepackage{markdown}
\usepackage{tabularx}
\usepackage[capitalise]{cleveref}
\begin{document}

\copyrightyear{2026}
\copyrightclause{Copyright for this paper by its authors.
  Use permitted under Creative Commons License Attribution 4.0
  International (CC BY 4.0).}

\conference{FRAME'26: Methodology First - Rethinking Research Assessment in RecSys Workshop, September 28, 2026, Minneapolis, Minnesota, USA}

\title{Calibrating Reproduced Claims in Recommender Systems}

\author{Alan Said}[orcid=0000-0002-2929-0529,email=alansaid@acm.org]
\address{University of Gothenburg, Gothenburg, Sweden}

\begin{abstract}
Reproduction studies can produce mixed outcomes. Reported values may differ while the ordering of the compared methods remains the same, a result may hold only under some experimental conditions, or a released implementation may fail to reproduce a result that the model can still reach. The terms repeatability, reproducibility, and replicability describe how a follow-up study relates to the original experiment, but not which parts of the original claim are supported by the new results. We introduce \emph{claim calibration} as a way of stating the strongest claim supported by a follow-up study, together with the conditions under which it holds and the parts that remain untested. We apply this perspective to five original--follow-up paper pairs from recommender-systems research. The cases show that agreement in numerical values, method rankings, statistical results, and overall conclusions does not always coincide, and that follow-up studies often support only part of the original claim. Based on these observations, we propose a Claim Evidence Profile for reporting the original claim, its scope, the reproduction target, the reported results, the calibrated claim, and the parts of the original claim that remain unresolved.
\end{abstract}

\maketitle

\section{Introduction}
\label{sec:introduction}

Repeatability, reproducibility, and replicability describe how a follow-up study relates to an original experiment. The terminology is not used consistently across research communities~\cite{goodman2016WhatDoesResearch}, so we use the current ACM terminology for these distinctions.\footnote{\url{https://www.acm.org/publications/policies/artifact-review-and-badging-current}} These terms are useful, but they do not specify what part of the original result was reproduced or what scientific conclusion can be drawn from the follow-up results. A study may recover the reported numerical values without testing whether the result holds under another setup. Another may obtain different values while preserving the ordering of the compared methods or the overall conclusion. A follow-up study may also introduce a stronger baseline or change part of the evaluation, thereby testing a narrower or different part of the original claim.

This distinction is particularly relevant in recommender systems, where results depend on choices such as baselines, sampling strategies, implementations, training setups, and evaluation measures~\cite{saidReproducibilityRecommenderSystems2026,belloginImprovingAccountabilityRecommender2021}. Changes to these choices can affect numerical values, method rankings, or the conditions under which a conclusion holds. Treating a reproduction study as a single success or failure can hide these differences.

We refer to the process of making this distinction explicit as \emph{claim calibration}. Claim calibration identifies the strongest claim supported by a follow-up study, together with the conditions under which it holds and the parts of the original claim that remain untested. We examine five original--follow-up paper pairs from a recent survey of reproducibility work in recommender systems~\cite{saidReproducibilityRecommenderSystems2026}. For each pair, we identify the main claim addressed by the follow-up study, how the study relates to the original setup, the results it reports, and the conclusion those results support. Based on this analysis, we propose a \textit{Claim Evidence Profile} for reporting what was tested, what was found, what remains supported, and what remains unresolved.

\section{Analysis of Reproduction Studies}
\label{sec:analysis}
This section applies claim calibration to five original--follow-up paper pairs from recommender-systems research. We do not try to assign one overall outcome to each follow-up study. Instead, we ask what part of the original claim was tested and what the reported results support.

\subsection{Case Selection}
\label{sec:case-selection}

We use the first five paper pairs listed in Table~9 of the reproducibility survey that motivated this study~\cite{saidReproducibilityRecommenderSystems2026}. These pairs are reproduced in \cref{tbl:deltas}. They cover collaborative filtering, sampled evaluation, conversational recommendation, evaluation beyond accuracy, and sequential recommendation. Their order was fixed before the claim-level analysis, so the selection was not based on whether the follow-up study confirmed or challenged the original result.

\begin{table}[tb]
\centering
\caption{The first five original--follow-up paper pairs from Table~9 in \cite{saidReproducibilityRecommenderSystems2026}. The last three columns show the number of algorithms, datasets, and metrics removed from (-) or added to (+) the original setup.}
\label{tbl:deltas}
\begin{tabular}{rcl rcl ccc}
\toprule
\multicolumn{3}{c}{Original} &
\multicolumn{3}{c}{Follow-up} &
\multicolumn{3}{c}{Changes} \\
Paper & Venue & Year & Paper & Venue & Year &
Algorithms & Datasets & Metrics \\
\cmidrule(r){1-3}
\cmidrule(lr){4-6}
\cmidrule(lr){7-9}
\cite{heNeuralCollaborativeFiltering2017}
& WWW & 2017
& \cite{rendleNeuralCollaborativeFiltering2020}
& RecSys
& 2020 & -2, +1 & -0, +0 & -0, +1
\\
\cite{kricheneSampledMetricsItem2020}
& KDD & 2020
& \cite{dallmannCaseStudySampling2021}
& RecSys
& 2021 & -2, +4 & -0, +4 & -5, +1
\\
\cite{chenKnowledgeBasedRecommenderDialog2019}
& EMNLP & 2019
& \cite{manzoorGenerationbasedVsRetrievalbased2021}
& RecSys
& 2021 & -2, +2 & -0, +0 & -4, +1
\\
\cite{rendleNeuralCollaborativeFiltering2020}
& RecSys & 2020
& \cite{anelliReenvisioningComparisonNeural2021}
& RecSys
& 2021 & -0, +3 & -0, +0 & -1, +15
\\
\cite{sunBERT4RecSequentialRecommendation2019}
& CIKM & 2019
& \cite{petrovSystematicReviewReplicability2022}
& RecSys
& 2022 & -6, +2 & -0, +0 & -1, +2
\\
\bottomrule
\end{tabular}
\end{table}

The five cases are not intended to represent all reproducibility work in recommender systems. They provide a small set of follow-up studies that differ in what they reuse, change, and test. The unit of analysis is a \emph{claim--evidence pair}: one main empirical claim from the original paper and the follow-up results used as evidence about that claim.

\subsection{Analytical Procedure}
\label{sec:analytical-procedure}

For each paper pair, we examined the original, the follow-up, and the public repositories when available. We recorded the main claim addressed by the follow-up study, the conditions attached to that claim, the parts of the experiment that were kept or changed, and the results reported. Experiments were not rerun; the analysis was based on the information reported in the papers and, where relevant, the accompanying public artifacts.

We preferred claims that were stated explicitly in the original paper. When a claim had to be inferred from the reported results or discussion, it was marked as implied. Results that could not be compared directly, e.g., because the measurement or evaluation procedure had changed, was treated as not comparable rather than as disagreement. Parts of the original claim that were not examined by the follow-up
study were recorded as not tested. 

Some of the follow-up studies reuse parts of the original setup while adding new implementations, models, or evaluation procedures. \cref{tab:cases} describes the relationship between the studies rather than assigning each pair to a single category. For each pair, we wrote a calibrated claim that states what the results support, under which conditions, and which parts of the original claim remain unresolved.

The complete coding records for all five cases are available as supplementary material.\footnote{\label{foot:material}\url{https://github.com/alansaid/calibratingclaims}} They include the selected original claim and its location, the target of the follow-up study, the reported results used in the analysis and their locations, the calibrated claim, and unresolved scope.

\begin{table}
\centering
\small
\caption{Summary of the five analyzed claim--evidence pairs.}
\label{tab:cases}
\begin{tabularx}{\textwidth}{
    >{\raggedright\arraybackslash}p{0.04\textwidth}
    >{\raggedright\arraybackslash}p{0.17\textwidth}
    >{\raggedright\arraybackslash}X
    >{\raggedright\arraybackslash}p{0.24\textwidth}
    >{\raggedright\arraybackslash}p{0.22\textwidth}}
\toprule
Case &
Study relationship &
Claim examined &
Reported Results &
Calibrated claim \\
\midrule
\cite{heNeuralCollaborativeFiltering2017} \&
\cite{rendleNeuralCollaborativeFiltering2020}
&
Reassessment using the original evaluation artifacts
&
Whether MLP-based similarities and NeuMF outperform
dot-product MF
&
A newly tuned MF baseline outperformed the reported NCF results
using the same datasets and evaluation protocol
&
The original experiments do not show that NCF is better than a
properly tuned MF baseline
\\
\midrule
\cite{kricheneSampledMetricsItem2020} \&
\cite{dallmannCaseStudySampling2021}
&
Independent test and extension with new models and datasets
&
Whether sampled ranking metrics preserve the model ordering
obtained through full-item evaluation
&
Uniform sampling produced different model rankings for four
sequential recommenders on five datasets; popularity sampling
showed the same problem
&
The sampling problem is supported for sequential recommenders
and popularity sampling, but the proposed corrections were not tested
\\
\midrule
\cite{chenKnowledgeBasedRecommenderDialog2019} \&
\cite{manzoorGenerationbasedVsRetrievalbased2021}
&
Reproduction with a different user- study setup and a new baseline
&
Whether KBRD produces responses of higher perceived quality
than DeepCRS
&
KBRD was again rated above DeepCRS, but a new retrieval-based
system was rated above both
&
KBRD's advantage over DeepCRS is supported, but the study does
not show that KBRD is better than other conversational approaches
\\
\midrule
\cite{rendleNeuralCollaborativeFiltering2020} \&
\cite{anelliReenvisioningComparisonNeural2021}
&
Independent reimplementation with additional evaluation measures
&
Whether a properly configured MF model outperforms NeuMF
under the original evaluation setup
&
The MF values were close to the earlier results, the NeuMF values
differed more, and MF still ranked above NeuMF; the additional
measures showed trade-offs between the models
&
MF remains the stronger model for accuracy in this setup, but not
for every evaluation measure
\\
\midrule
\cite{sunBERT4RecSequentialRecommendation2019} \&
\cite{petrovSystematicReviewReplicability2022}
&
Evaluation of the original implementation, an independent
implementation, and published comparisons
&
Whether BERT4Rec consistently outperforms SASRec and other
sequential recommendation models
&
The default settings did not recover the reported results; longer
training and an independent implementation recovered most of them,
while published BERT4Rec--SASRec comparisons were mixed
&
BERT4Rec can reach the reported effectiveness when sufficiently
trained, but it does not consistently outperform SASRec across
implementations and studies
\\
\bottomrule
\end{tabularx}
\end{table}

\subsection{Cross-Case Findings}
\label{sec:cross-case-findings}

\Cref{tab:cases} summarizes the five claim--evidence pairs. We focus on three differences across the cases.

First, a follow-up study can test an original claim without rebuilding the full original experiment. \citet{rendleNeuralCollaborativeFiltering2020} used the evaluation setup of \citet{heNeuralCollaborativeFiltering2017} but added a stronger matrix-factorization baseline. They did not regenerate all of the NCF results. The main question was instead whether the original comparison still held when a better-tuned baseline was included. Similarly, \citet{dallmannCaseStudySampling2021} did not repeat the exact experiments of \citet{kricheneSampledMetricsItem2020}. They tested the same claim about sampled metrics using other models and datasets. Their results support the claim in these new settings, but they do not test the corrected metrics proposed in the original paper.

Second, different parts of a result can agree or disagree. \citet{anelliReenvisioningComparisonNeural2021} obtained different
values from the earlier study, particularly for NeuMF, but MF still ranked above NeuMF \cite{rendleNeuralCollaborativeFiltering2020}. The main accuracy comparison stayed the same even though the values did not match exactly. When more evaluation measures were added, however, MF was not better on every measure. In the conversational recommendation case, \citet{manzoorGenerationbasedVsRetrievalbased2021} used a different rating scale, a different study design, and a different group of participants \cite{chenKnowledgeBasedRecommenderDialog2019}. The scores cannot be compared directly, but KBRD was still rated above DeepCRS. The new retrieval-based system was rated above both, so the follow-up study supports the KBRD--DeepCRS comparison but not a broader claim that KBRD is the best approach.

Third, the default implementation may fail to reproduce a result even when the model can reach it. \citet{petrovSystematicReviewReplicability2022} found that the default BERT4Rec settings did not recover the results reported by \citet{sunBERT4RecSequentialRecommendation2019}. With longer training, the original implementation reached the reported performance in the examined setting. An independent implementation also recovered most, but not all, of the results. Their review of later papers further showed that BERT4Rec did not consistently outperform SASRec.

Together, the cases show why one overall reproduction label is often not enough. A follow-up study may reproduce values, keep a ranking, support only one part of a claim, or show that a result depends on a particular setup. The Claim Evidence Profile in the next section is intended to make these differences easier to report.

\section{The Claim Evidence Profile}
\label{sec:profile}

The cases in Section~\ref{sec:analysis} show that one reproduction label is often not enough. We propose the \emph{Claim Evidence Profile} as a compact way to report what a follow-up study tested and what its results support. The profile is written for one claim at a time, since a study may support some claims while leaving others untested. Complete profiles for the five cases analyzed in this paper are included in the supplementary material.\footref{foot:material}

The profile is inspired by structured documentation approaches such as Model Cards \cite{mitchellModelCardsModel2019}, Datasheets for Datasets  \cite{gebruDatasheetsDatasets2021}, and Impact Assessment Cards \cite{boguckaImpactAssessmentCard2025}. 
Structured reporting has also been used to improve reproducibility in machine learning, for example through the NeurIPS reproducibility checklist~\cite{pineau2021ImprovingReproducibilityMachine}. Such checklists focus primarily on whether the information needed to understand and reproduce an experiment is reported.  The Claim Evidence Profile follows the same general idea, but documents a different object. Rather than describing a model, dataset, or application, it records the relationship between an empirical claim and the results reported when that claim is revisited.
Its purpose is not general documentation, but to make the step from reported results to the claim supported by those results explicit.

\begin{table}[t]
\centering
\small
\caption{Fields of the Claim Evidence Profile.}
\label{tab:profile-fields}
\begin{tabularx}{\textwidth}{
    >{\raggedright\arraybackslash}p{0.18\textwidth}
    >{\raggedright\arraybackslash}X}
\toprule
Field & Description \\
\midrule
Original claim &
The conclusion stated or implied by the original study. \\
Scope &
The task, data, models, baselines, protocol, metrics, and other
conditions attached to the claim. \\
Study relationship &
How the follow-up study relates to the original setup and artifacts. \\
Reproduction target &
The result, comparison, or conclusion tested by the follow-up study. \\
Reported results &
The results reported by the follow-up study that bear on the claim.\\
Calibrated claim &
The strongest conclusion supported by the reported results, taking their stated limitations into account. \\
Unresolved scope &
The parts of the original claim and the conditions that were not
tested. \\
\bottomrule
\end{tabularx}
\end{table}

\subsection{Profile Structure}
\label{sec:profile-structure}

The profile contains seven fields, listed in \cref{tab:profile-fields}. The original claim should follow the wording of the source paper as closely as possible. The scope lists the conditions attached to that claim. The reproduction target states which part of the claim the follow-up study tested.

The relationship between the studies and the results of the follow-up study are reported separately. An independent implementation may produce different values while preserving the same model ranking. A study using the original artifacts may recover the reported values without testing whether the result also holds under another setup.
The calibrated claim states what the reported results support and under which conditions. The profile also separates evidence against a claim from a claim that was not tested. A missing test should not be interpreted as the claim being wrong.

The profile records the results reported by a follow-up study and how they bear on the original claim. It does not, by itself, establish the methodological quality of that evidence. The calibrated claim should also reflect limitations in the evidence used to support it. These may include, for example, insufficiently tuned baselines, inappropriate statistical analysis, data leakage, small samples, or limitations of the evaluation procedure. 

\begin{table}[h]
\centering
\small
\caption{Claim Evidence Profile for the NCF--MF case \cite{heNeuralCollaborativeFiltering2017, rendleNeuralCollaborativeFiltering2020}.}
\label{tab:profile-example}
\begin{tabularx}{\textwidth}{
    >{\raggedright\arraybackslash}p{0.18\textwidth}
    >{\raggedright\arraybackslash}X}
\toprule
Field & Entry \\
\midrule
Original claim &
Learned MLP similarities and NeuMF outperform established collaborative-filtering baselines. \\
Scope & 
MovieLens-1M and Pinterest, sampled-candidate evaluation, HR@10 and NDCG@10, and the implementations used in the study. \\
Study relationship &
The original evaluation artifacts were reused, while a new MF baseline was implemented and tuned independently. \\
Reproduction target &
Whether the reported NCF results show that NCF outperforms a properly configured dot-product model. \\
Reported results & 
The original NCF values were not regenerated. The new MF baseline generally outperformed the reported MLP and NeuMF results. \\
Calibrated claim &
Under the original datasets and evaluation setup, the reported experiments do not show that an MLP-based similarity is better than a properly tuned dot-product baseline. \\
Unresolved scope &
Other datasets, full-catalogue evaluation, alternative losses, implementations, and tuning procedures. \\
\bottomrule
\end{tabularx}
\end{table}

\subsection{Example}
\label{sec:profile-example}

\Cref{tab:profile-example} applies the profile to the comparison between Neural Collaborative Filtering \cite{heNeuralCollaborativeFiltering2017} and Matrix Factorization \cite{rendleNeuralCollaborativeFiltering2020}. \citet{rendleNeuralCollaborativeFiltering2020} reused the original datasets and evaluation setup, but added a newly implemented and tuned matrix-factorization baseline. The study tested whether the original comparison still held when this baseline was included. It did not regenerate all of the original NCF results.

The profile does not assign a score to either paper. It shows the original claim, the part tested by the follow-up study, the reported results, and the claim those results support. It can be used by authors when reporting a reproduction study, by reviewers when assessing its conclusions, and by readers when comparing studies.

\section{Discussion and Conclusion}
\label{sec:discussion}

The five cases in \cref{tab:cases} show that reproduction is easier to interpret at the level of individual claims than at the level of whole papers. A follow-up study may recover the reported values, preserve the ranking of the compared methods, support only part of the original conclusion, or show that the result depends on a particular setup. Assigning one outcome to the whole paper hides these differences.

Disagreement between two studies can also mean different things. Different numerical values may still lead to the same ranking and the same conclusion. A matching result, on the other hand, may say little about whether the result holds with another baseline, dataset, implementation, or evaluation protocol. A follow-up study may also provide relevant evidence without directly rerunning the original experiment, for example by introducing a stronger baseline or testing the same claim in another setting. These cases should not be reduced to the same success or failure label.

Claim calibration does not replace the existing terminology. Repeatability, reproducibility, and replicability describe how the follow-up study relates to the original study. The Claim Evidence Profile describes what was tested and what the reported results support. Both are needed. An independent implementation may produce different values but the same ranking, while an original implementation may reproduce the reported result only after its default settings are changed.

The profile is intended as a reporting aid rather than a score. Its purpose is to make the connection between evidence and conclusion easier to inspect. Authors can use it to state which claim they tested, which parts of the original setup they kept or changed, and what they found. Reviewers can use it to check whether the conclusion follows from the evidence. Readers can use it to see whether two studies disagree about the same claim or have tested different parts of it. The profile also does not replace an assessment of the quality of the underlying experiment. A well-documented piece of evidence can still be weak evidence, and this should limit the claim drawn from it.

There is, however, a limit to how detailed such reporting can be. An empirical paper may contain many claims, and each claim may depend on a large number of experimental choices. Completing a profile for every result would quickly become impractical. The profile is better suited to the claims that are central to the paper or directly addressed by the follow-up study. The aim is not to document every implementation detail, but to make clear which conclusion is being examined and what the new results change.

The distinction between missing evidence and evidence against a claim is also important. A follow-up study may test one part of an original contribution while leaving other parts untouched. Those untested parts should not be described as reproduced, but they should not be described as disproved either. This matters when later work is used to summarize the status of an earlier result, since a narrow reproduction attempt can otherwise be read as a judgment on a much broader set of claims.

This study is limited to five paper pairs selected from a larger survey. The cases are illustrative rather than representative, the coding was carried out by one researcher, and the analysis is based on the papers and available artifacts rather than rerun experiments. The calibrated claims are interpretations of the reported results, and another researcher may draw the boundary between supported and unsupported parts of a claim differently. To make these interpretations inspectable, we provide the complete coding records, source locations, and coding rules as supplementary material.\footref{foot:material}

Future work should include more cases, multiple coders, and an assessment of coding agreement. It would also be useful to study how authors and reviewers use the profile in practice, including whether it helps identify overstated conclusions and whether it is practical to complete as part of a reproduction study.

The seven fields proposed here should be considered a first version of the profile; a broader analysis of reproduction studies may reveal additional distinctions that should be represented.

Reproducing a result can mean different things. It may refer to code execution, numerical values, method rankings, statistical results, overall conclusions, or results under changed conditions. Reporting which of these was tested and supported gives a clearer account of what a follow-up study has shown. Claim calibration provides one way of making that account more explicit.

\section*{Declaration on Generative AI}
During the preparation of this work, the author(s) used ChatGPT and Grammarly for Grammar and
spelling check and rephrasing. After using these tool(s)/service(s), the author(s) reviewed and edited the content as needed and take(s) full responsibility for the publication’s content.

\bibliography{sample}

\end{document}